\documentclass[doublecol]{epl2} 
\RequirePackage{amssymb} 
\usepackage[colorlinks = true, linkcolor = black, citecolor = black, urlcolor = blue]{hyperref}
\usepackage{breakurl}

   \title{Complex dynamics of knotted filaments in shear flow}
\shorttitle{Complex dynamics of knotted filaments in shear flow} %Insert here a short version of the title if it exceeds 70 characters

\author{R. Matthews \and A.A. Louis  \and J.M. Yeomans}
\shortauthor{R. Matthews \etal}

\institute{                    
  \inst{1} Rudolf Peierls Centre for Theoretical Physics, 1 Keble Road, Oxford 0X1 3NP, England\\
  }


\abstract{Coarse-grained simulations are used to demonstrate that knotted filaments in shear flow at zero Reynolds number exhibit remarkably rich dynamic behaviour.  For stiff filaments that are weakly deformed by the shear forces, the knotted filaments rotate like rigid objects in the flow. But away from this regime the interplay between between shear forces and the flexibility of the filament leads to intricate regular and chaotic modes of motion that can be divided into distinct families. The set of  accessible mode families depends to first order on a dimensionless number that relates the filament length, the elastic modulus, the friction per unit length and the shear rate.}
\begin{document}
\bibliographystyle{eplbib}
 
\maketitle

The interaction between a shear gradient and suspended objects can generate fascinating dynamical behaviour.
 G. Jeffrey~\cite{jeffrey} showed in 1922 that rigid bodies trace out complex periodic orbits that depend in detail on their shapes. For deformable objects~\cite{ralliston84} even richer behaviour is possible. Red blood cells, for example, change shape with increasing shear rate~\cite{schmid69}.  Furthermore, the coupling to shear can lead to either unstable tumbling or to a steady state mode where the cells remain at a fixed angle to the flow while their outer membrane rotates like the treading of a tank~\cite{noguchi}. A similar crossover to tank treading is predicted for star polymers~\cite{ripoll}. Experimental advances in single molecule techniques have made it possible to directly observe the stretching and tumbling behaviour of individual DNA molecules~\cite{smith}. This work inspired a great deal of theoretical research on the way that the polymer flexibility, shear and Brownian noise interact~\cite{shaqfeh}. Several decades earlier it had been shown that  filaments in the non-Brownian regime exhibit at least  five distinguishable regimes of motion as stiffness,  length, and shear rate are varied~\cite{forgacs}. These results are still the subject of active investigation by theorists~\cite{lindstrom}. The crossover between the Brownian and non-Brownian regimes has also been recently considered~\cite{kobayashi}.

In this paper we use coarse-grained computer simulations to study the behaviour of knotted non-Brownian  filaments in shear flow. Knots are a generic possibility for any long elastic objects and occur naturally  in biologically active  DNA~\cite{watt,arsuaga}. In the limit of strong bending modulus $A$ or weak shear rate $\dot{\gamma}$ the knotted filament will take on its equilibrium shape~\cite{gallotti}, and rotate in a manner similar to that first predicted by Jeffrey~\cite{jeffrey}.  Chiral knots should also migrate in the vorticity direction~\cite{kim,watari}, a hydrodynamic effect that has recently been observed for other objects including helical bacteria~\cite{marcos}.   

%-- FIGURE 1 ------
\begin{figure*}[!htpb]
\begin{center}
\includegraphics[scale=0.7]{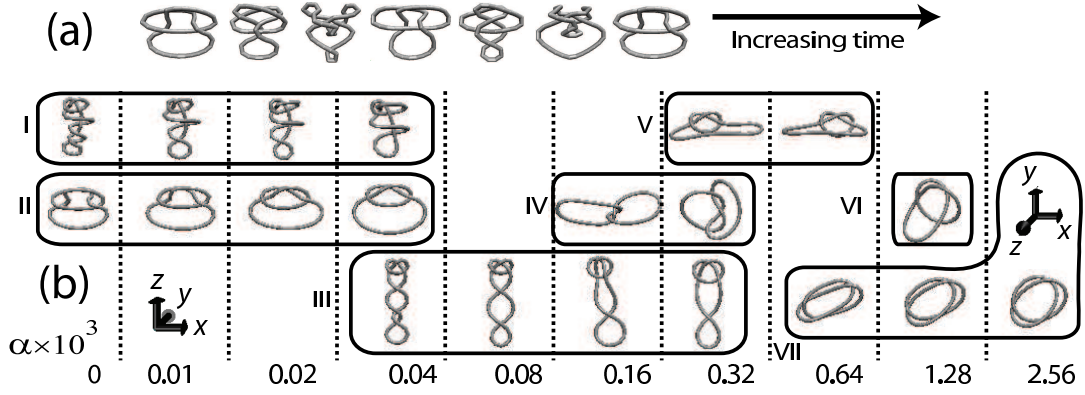}
\caption{\label{fig:mode_configs} Predicted modes of behaviour as a function of the knot deformation number $\alpha$. $x$ is the flow direction, $y$ is the direction of the shear gradient,   and $z$ is the vorticity direction. All configurations are for filaments with a $3_1(-)$ trefoil knot (a) Time series for filament in a II mode at $\alpha= 0$. Time increases from left to right. The knot rotates clockwise around the $z$ axis.  The series spans approximately one period, the time between successive configurations is $1028.7t_0$ and migration is in the  $-z$-direction. (b) Example configurations for different mode families over a range of $\alpha$. Note that the orientation is different for the VII mode. Full animations of the modes are available online~\cite{animations} (see Table~\ref{tab:animations} in the appendix).}
\end{center}
\end{figure*}

The focus of this paper, however, is  what happens for stronger shear forces and/or for more flexible filaments, i.e.\ the regime where the knots can be tightened by the flow. To estimate where this crossover  occurs we consider the following argument: in the limit of small local bond deformations, the bending energy of a knot of length $L$ scales as $\sim A/L$ so there is a force opposing knot tightening $\sim A/L^2$.  Neglecting the logarithmic factor~\cite{cox}, the drag on a slender filament in Stokes flow $\sim\eta L$. To tighten the knot, strands which are close to each other must be moved in opposite directions The typical velocity difference will be $\sim\sigma\dot{\gamma}$, where $\sigma$ is the filament width. Combining this with the drag and comparing to the bending force results in a dimensionless knot deformation number, $\alpha = A/\eta\sigma\dot{\gamma} L^3$. (At the crossover, the knot length $\approx$ total filament length, so in calculating $\alpha$ we take $L$ to be the total filament length.) This resembles the sperm number used with microscopic swimmers~\cite{lowe} but instead of determining when the filament as a whole may be deformed, it indicates when the knot will be tightened. For larger $\alpha$ we expect the stiff knot regime, but for lower $\alpha$ the shear should cause significant deformation.
 
Indeed, as illustrated in Fig.~\ref{fig:mode_configs} for the case of the simple trefoil knot, lowering $\alpha$ leads to a crossover from knots that remain close to their equilibrium shape, to a regime of surprisingly rich dynamical behaviour where the whole filament exhibits intricate  shape oscillations in time. These orbits may be grouped into a few distinct families comprising very similar types of motion (modes). Some modes show regular, and others chaotic, motion. Different modes show distinct rates and directions of drift along the vorticity axis. A few families are accessible at each $\alpha$, and the  knot typically falls into one type of motion depending on initial conditions. Changing $\alpha$ therefore changes the modes that are accessible, as well as the probability that a certain mode is selected. In the appendix we list details of videos, accessible online~\cite{animations}, which illustrate the modes.

In the rest of this paper we describe how we simulate the knotted filaments and analyse the ensuing results. We apply a coarse-grained bead-spring model~\cite{yamamoto}. The interaction potential between beads is~\cite{grest}: \begin{eqnarray}
\nonumber V&=&-\kappa\sum_{i}\frac{\vec{r}_{i,i+1}\cdot\vec{r}_{i-1,i}}{r_{i,i+1}r_{i-1,i}}  + \sum_{j>i}\sum_{i}H\left[ 2^{\frac{1}{6}}\sigma -  r_{ij}\right]
\\\nonumber&\times&4\epsilon\left[\left(\frac{\sigma}{r_{ij}}\right)^{12}-\left(\frac{\sigma}{r_{ij}}\right)^{6}+\frac{1}{4}\right]
\\&-&\frac{k 
R_0^2}{2}\sum_{i}\ln\left[1-\left(\frac{r_{i,i+1}}{R_0}\right)^2\right]
\label{potential}
\end{eqnarray} where $\vec{r}_i$ is the $i^{th}$ bead position and $\vec{r}_{ij} = \vec{r}_i - \vec{r}_j$. The first term in Eq.~(\ref{potential}) allows the flexibility of the filaments to be varied by changing $\kappa$, the bending energy. The second term is excluded volume: $H$ is the Heaviside step function which truncates the Lennard-Jones potential to be purely repulsive, $\sigma$ and $\epsilon$ are the length and energy scales respectively. The third term is a FENE spring potential. We choose $k = 30{\epsilon}/{\sigma^2}$ and $R_0 = 1.5\sigma$.

We update bead positions using the Euler method~\cite{watari}: 
\begin{equation}
\vec{r}_i(t+\Delta t) = \vec{r}_i(t) + \left[\vec{v}(\vec{r}_i)+\sum_{j} 
\mathcal{H}_{ij}\cdot\vec{f}_j\right]\Delta t
\label{timestep}
\end{equation} 
where $\vec{v}(\vec{r}_i)$ is the applied flow, $\vec{v}=\dot{\gamma}y\vec{\hat{x}}$ and  $\vec{f}_i$ is the force on the $i^{th}$ bead which results from the potential~(\ref{potential}). Assuming zero Reynolds number, the hydrodynamic interactions, $\mathcal{H}_{ij}$, can be approximated by the Rotne-Prager-Yamakawa interaction tensor~\cite{rotne} with viscosity $\eta$, and taking the hydrodynamic radius as $\sigma / 2$.
The natural input units for the simulation are $\sigma$, $\eta$ and $k$, from which a natural time unit of $t_0={\sigma\eta}/{k}$ follows. We used $\dot{\gamma}=(150 t_0)^{-1} $, and  verified that the stretching of individual bonds relaxes much more quickly than the characteristic time for shear induced motion and so should not influence the dynamics~\cite{powers}.

We simulated single filament rings with one knot, using chains  $N=50$ beads (unless otherwise stated). We mainly study the chiral trefoil knot, standardly denoted $3_1$~\cite{orlandini}, choosing, unless otherwise stated, the left-handed enantiomer ($3_1(-)$), not the right-handed ($3_1(+)$).

Filaments were given a knotted configuration and then equilibrated for $6\times10^5t_0$ at finite temperature with no shear to generate random starting points. A $6\times10^6t_0$ initialisation period was allowed before data was recorded for $1.5\times10^6t_0$. The relatively long initialisation was chosen to avoid transients. We ran 50 simulations for most parameter sets. The majority of runs exhibited the same mode of motion throughout the observation time, but for a small number (less than 1\%) of runs a slightly longer initialisation was necessary. The long transients may caused by the system being close to a parameter value at which a mode appears/disappears~\cite{grebogi}. Eq.~(\ref{timestep}) was typically integrated using $\Delta t = 10^{-2}$ although for some parameter choices, for example high $\kappa$, it was necessary to reduce this for numerical stability. We checked that using $\Delta t = 10^{-3}$ gave equivalent results.

Mode families were identified by visual inspection of their motion (see e.g.\ the videos in~\cite{animations}) and by measuring their drift velocity in the vorticity direction, which distinguished well between different modes. The identity was confirmed by two additional order parameters which measure the direction of maximum extension and the asymmetry under a $\pi$ rotation about $z$. Detailed definitions, as well as plots of average values for individual runs, are given in the appendix.

The elastic modulus of our bead-spring model  $A=\kappa \sigma$~\cite{lowe}, so that the knot deformation number defined earlier takes the form $\alpha = \kappa/\eta \dot{\gamma} L^3$.  
In Fig.~\ref{fig:mode_configs}  we consider 10 different values of $\kappa$ corresponding to $\alpha = 0 - 2.56\times10^{-3}$. For the largest $\alpha$, the filament remains in braid-like configurations~\cite{gallotti} that characterise family VII in Fig.~\ref{fig:mode_configs}(b). The motion is composed partly of rotation of the configuration and partly of tank-treading -- a particular point moves around the contour. As flexibility is increased ($\alpha$ is lowered) there is a change from modes which rotate in the $x$-$y$-plane to modes which have relatively large extensions in the $z$-direction. Interestingly, a similar shift was seen in experiment with linear filaments~\cite{forgacs}. The first family in which the knot is significantly tightened is V.

\begin{figure}
\includegraphics[scale=0.24]{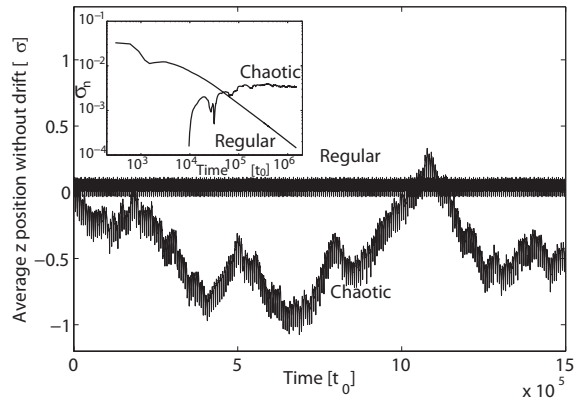}
\caption{\label{fig:reg_chaos_inset} Comparing regular and chaotic runs with modes in family II at $\alpha = 0$. The average $z$-position of the filament beads with the average drift subtracted is plotted as a function of time. The inset shows the estimate of the largest Lyapunov exponent as function of time for the two runs, plotted on a log-log scale.}
\end{figure}

Some mode families show both regular and chaotic modes, sometimes at the same $\alpha$, for example family II. Others showed only regular (VI), or only chaotic (IV), motion. To distinguish regular and chaotic modes, we calculated the largest Lyapunov exponent, $\sigma_1$~\cite{lichtenberg}:  A second system was created with the bead positions $\vec{r}_i$ each randomly displaced to $\vec{s}_i$ and constrained so that $d^2=\sum_{i}\mid\vec{r}_i-\vec{s}_i\mid^2/{\sigma^2}=1$. Both systems were integrated forward in time. After each  $1.5 t_0$,  $\vec{r}_i-\vec{s}_i$ were rescaled by changing $\vec{s}_i$ so as to make $d = 1$. The Lyapunov exponent is then given by~\cite{lichtenberg}: $\sigma_1 \equiv \mbox{lim}_{n \rightarrow \infty} \sigma_n = \mbox{lim}_{n \rightarrow \infty}(2/3nt_0)\sum_{i=1}^n \ln (d_j)$, where $d_j$ is the distance after the $j^{th}$ evolution. If the measured $\sigma_n$ tended to zero or a positive constant as a function of $n$ the motion was identified as regular or chaotic respectively.  This behaviour is illustrated in the inset of Fig.~\ref{fig:reg_chaos_inset}. 

Fig.~\ref{fig:reg_chaos_inset} compares the movement of the average bead position around its average drift in the $z$-direction for  two modes from family II at $\alpha = 0$, one regular and one chaotic. The regular mode simply oscillates. By contrast the chaotic mode, whilst showing oscillations, also displays larger movements. The power-spectrum of the curve for chaotic motion, shown in the appendix, exhibits a power-law decay with an exponent of about minus 2 ( $1.94 \pm 0.03$), suggesting a random walk around the average drift.
 
%----- FIGURE 2 -------

\begin{figure}
\includegraphics[scale=0.24]{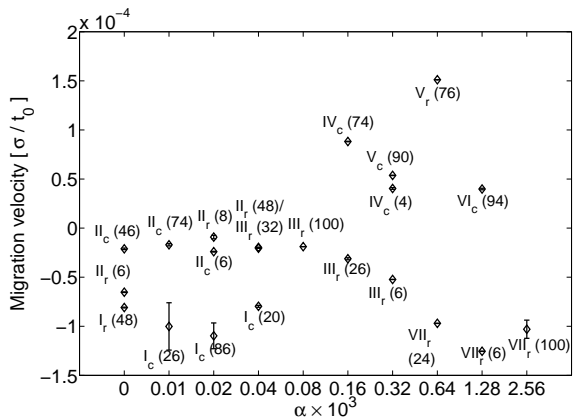}
\caption{\label{fig:mig_vels_fams} Migration velocity in the $z$-direction, averaged over all runs belonging to each family.  The labels indicate families and are subdivided into regular and chaotic modes, denoted by the subscripts  $r$ and $c$  respectively. Error bars show the standard deviations. The figures in brackets indicate the percentage of runs at each $\alpha$ that were observed to fall into each group.}
\end{figure}

Average migration velocities were calculated by a linear fits to the $z$-displacements of the centre of resistance. Fig.~\ref{fig:mig_vels_fams} plots the averages over all runs, grouped into families and then subdivided into regular and chaotic modes at each $\alpha$. Error bars indicate the spread of velocities observed. For most they are smaller than the data points. Fig.~\ref{fig:mig_vels_fams} also shows that at some $\alpha$ there exist modes that migrate in opposite directions. The period of rotation of regular modes varies with $\alpha$,  approximately in the range $2 - 6 \times 10^3 t_0$. The shortest periods were seen for the highest $\alpha$. For example at $\alpha = 0$ the average period of modes in the II family was $6220 \pm 40 t_0$ and that of family VII at $\alpha = 2.56 \times 10 ^{-3}$ was $2040 \pm 90 t_0$. Fig.~\ref{fig:mig_vels_fams}  also shows the percentage chance that a run with random initial conditions ends up in the particular mode family.

We also considered simulations for other knot types at $\alpha = 0$. For $3_1(+)$ the same modes are seen but the migration, and the orientation, is as expected, in the opposite $z$-direction. We see similar behaviour -- regular and chaotic modes with migration -- for more complex knots such as $4_1$ and $5_1$. For achiral $4_1$ the distribution of migration velocities is symmetric about zero: all migrating modes have a partner with opposite migration direction.  Depending on initial conditions, the $4_1$ knot may thus migrate in the $\pm z$-direction, but that the average migration velocity over many runs would be zero.

%--- FIGURE 3 -----
\begin{figure}
\includegraphics[scale=0.33]{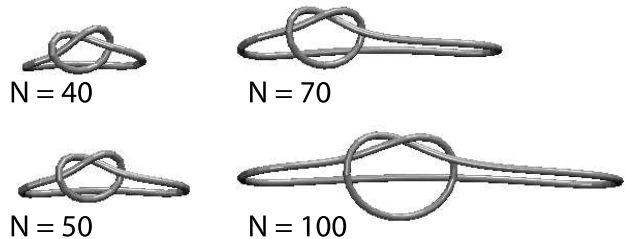}
\caption{\label{fig:vary_N_V_configs} Example configurations of the modes at $\alpha = 0.64 \times 10^{-3}$ for $N$ = 40, 50, 70 and 100. Full animations of the modes are available online~\cite{animations} (see Table~\ref{tab:animations} in the appendix).}
\end{figure}

Finally, we consider the sensitivity of our results to changes in parameters and changes in simulation details. It should be kept in mind that for these dynamical systems with behaviour that may depend sensitively on initial conditions, one would expect quantitative changes when simulation details are changed. The main thrust of our paper is qualitative, and so the most important tests will be whether the overall behaviour, i.e.\ the mode families, are robust to these changes.

Firstly we consider the effect of hydrodynamic interactions by setting $\mathcal{H}_{ij}=0$ for $i \neq j$ in Eq.~(\ref{timestep}).  We find, as expected, that the motion in the vorticity direction is a consequence of off-diagonal hydrodynamic interactions. Modes that resemble those of  families V and VII were seen but none with large $z$-extensions.

Secondly, linear filaments or unknotted rings with point force hydrodynamics simply align in the $x$-$z$ plane without access to different shear velocities. Simulations must  therefore explicitly take the finite thickness into account by  considering the torque on individual beads~\cite{lindstrom}. We tested this sensitivity by using algorithms that include the torque, and find that, in contrast to unknotted filaments, similar mode families are observed. The knot forces the system out of the plane so that it always accesses different shear velocities, and this dominates. 

Thirdly, we checked how the behaviour is affected by changing $\kappa$, $\eta$ and $\dot{\gamma}$ 
in such a way as to keep $\alpha$ fixed: in the absence discretisation effects, such changes of parameters should lead to descriptions of the same physical system and so the same behaviour is expected. We ran two sets of simulations where $\dot{\gamma}$ was reduced by a factor of 10 and either $\kappa$ was decreased or $\eta$ increased to compensate. We obtained very similar results. 

Finally, most of our results are for a fixed length filament with $N=50$ beads.  It is interesting to investigate how sensitive our results are to the length $L=N\sigma$.   For example,  for $N$ = 50 significant tightening first occurs in family V at $\alpha = 0.64\times 10^{-3}$. We ran additional simulations with $N$ = 40, 70 and 100 at $\alpha =0.64\times 10^{-3}$. In each case the majority of runs show a mode very similar to those in family V, see Fig.~\ref{fig:vary_N_V_configs}. The migration velocities are similar but decrease with $L$. The measured velocities are 1.60, 1.51, 1.35 and 1.27 $\times 10^{-4} \sigma / t_0$ respectively (in each case all runs in the mode had exactly the same velocity to the accuracy given.) At other $\alpha$s we checked the results at different $N$ were also qualitatively similar, although the agreement worsens at lower $\alpha$. For example at $\alpha = 1.28\times 10^{-3}$ modes like families VI and VII were seen for all $N$ but for $\alpha < 0.16\times10^{-3}$ we observed families for $N = 100$ that were qualitatively different to any seen for other $N$. In fact, we expect substantial differences at small $\alpha$ because the shape of the tighter knot is then fixed by the excluded volume of the chain and not just by the physics that enters into the derivation of $\alpha$.  In that regime, for fixed filament thickness, we expect the influence of the knot to become progressively smaller as $L/\sigma \rightarrow \infty$.  While it would be interesting to explore these effects further, at fixed $\dot{\gamma}$, changing $N$ while fixing $\alpha$ means that $\kappa$ must be increased as $N^3$ and the integration timestep correspondingly decreased for stability. Combining this with $N^2$ time for calculation of $\mathcal{H}_{ij}$ gives a prohibitive $\sim N^5$ scaling of simulation time.

To summarise, we have demonstrated that knotted filaments in shear exhibit a rich dynamical behaviour with modes which can be divided into families. Some families have both regular and chaotic modes. Mode families migrate in different directions along the vorticity axis. The crossover from a stiff knot to the regime where multiple modes are possible can be described by a dimensionless number. In future work it may be interesting to consider more sophisticated treatments of the hydrodynamics that include effects such as lubrication. It may also be interesting to consider the effect of noise: Initial simulations suggest that fluctuations may alter the stability of modes leading to a variation or even flipping of migration velocity as function of noise strength.

Experimentally, this behaviour would be most easily observable with macroscopic filaments in highly viscous solutions~\cite{forgacs}.  However, it may also be visible for  DNA. For example,  the P4 phage genome (common in knotting experiments~\cite{arsuaga}) is about $77$ thermal persistence lengths long We estimate a crossover ($\alpha = 10^{-3}$) at  $\dot{\gamma} \approx 2\times10^3 s^{-1}$ in water.  The Weissenberg number $\approx 10$ so shear should be reasonably strong compared to thermal effects.

\section{Appendix}

We present a range of animations of modes from the families described in the main text.  The filenames, along with additional information, are listed in Table~\ref{tab:animations}. All animations are of duration $9000 t_0$ and of simulations at a shear rate of $\dot{\gamma}=(150 t_0)^{-1}$ with a $3_1(-)$ knot. The green sections of the filaments are markers to allow the motion to be more easily followed.

\begin{table}[!htpb]
\caption{\label{tab:animations} Examples of animations of modes belonging to the various families described in the main text. Files may be accessed online~\cite{animations}. Animations were created using VMD~\cite{humphrey}.}
\begin{center}
\begin{tabular}{l|c|c|c|r} \hline

Filename  & Mode  &  Regular/ & $N$ & $\alpha \times 10^3$  \\
                   & Family &  Chaotic  &          &                                              \\ \hline

 \href{http://iopscience.iop.org/0295-5075/92/3/34003/media/fam1r.mpg}{{\color{blue}\underline{fam1r.mpg}}} & I   & r & 50 & 0 \\ 
 \href{http://iopscience.iop.org/0295-5075/92/3/34003/media/fam2r.mpg}{{\color{blue}\underline{fam2r.mpg}}} & II  & r & 50 & 0 \\ 
 \href{http://iopscience.iop.org/0295-5075/92/3/34003/media/fam2c.mpg}{{\color{blue}\underline{fam2c.mpg}}} & II  & c & 50 &0 \\ 
 \href{http://iopscience.iop.org/0295-5075/92/3/34003/media/am3r.mpg}{{\color{blue}\underline{fam3r.mpg}}}& III & r &  50 &0.04 \\   
 \href{http://iopscience.iop.org/0295-5075/92/3/34003/media/fam3r2.mpg}{{\color{blue}\underline{fam3r2.mpg}}} & III & r & 50 &0.16 \\ 
 \href{http://iopscience.iop.org/0295-5075/92/3/34003/media/fam4c.mpg}{{\color{blue}\underline{fam4c.mpg}}}& IV& c & 50 &0.16 \\ 
 \href{http://iopscience.iop.org/0295-5075/92/3/34003/media/fam5r.mpg}{{\color{blue}\underline{fam5r.mpg}}} & V & r & 50 &0.64 \\
 \href{http://iopscience.iop.org/0295-5075/92/3/34003/media/fam6c.mpg}{{\color{blue}\underline{fam6c.mpg}}} & VI& c & 50 &1.28 \\  
 \href{http://iopscience.iop.org/0295-5075/92/3/34003/media/fam7r.mpg}{{\color{blue}\underline{fam7r.mpg}}}& VII& r & 50 &0.64 \\ 
 \href{http://iopscience.iop.org/0295-5075/92/3/34003/media/fam5rN40.mpg}{{\color{blue}\underline{fam5rN40.mpg}}} & V & r & 40 &0.64  \\ 
 \href{http://iopscience.iop.org/0295-5075/92/3/34003/media/fam5rN70.mpg}{{\color{blue}\underline{fam5rN70.mpg}}} & V & r & 70 & 0.64  \\ 
 \href{http://iopscience.iop.org/0295-5075/92/3/34003/media/fam5rN100.mpg}{{\color{blue}\underline{fam5rN100.mpg}}} & V & r & 100 &0.64  \\ 

 \hline

\end{tabular}
\end{center}
\end{table}

\begin{figure}
\begin{center}
\includegraphics[scale=0.35]{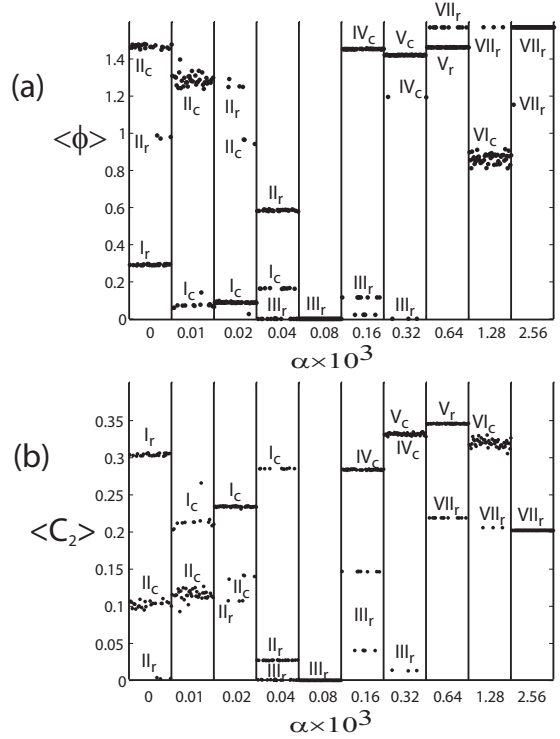}
\caption{\label{fig:N50_modes_sup} The values of the order parameters, averaged over single runs for $N=50$ filaments at different values of $\alpha$. (a) The angle of the direction of maximum extension to the $z$-axis, $\phi$. The averages of $\phi$ for each run are plotted for a given $\alpha$ in an arbitrary order. The labels indicate the modes to which the different groups of points correspond. (b) The same as (a) but for $C_2$, an order parameter to detect two-fold symmetry about the $z$-axis. Lower values indicate more symmetric configurations. It should be emphasised that all the points within two consecutive vertical lines correspond to different runs at the same $\alpha$ -- the positions along the $x$-axis within each section are irrelevant.
}
\end{center}
\end{figure}

\begin{figure}
\begin{center}
\includegraphics[scale=0.25]{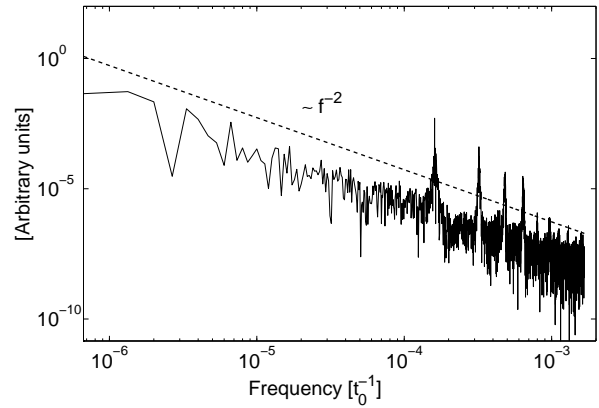}
\caption{\label{fig:chaotic_pow_spec} Power spectrum calculated by discrete Fourier transform of the displacement about the average drift for the chaotic mode plotted in Fig.~\ref{fig:reg_chaos_inset}. The dashed line has a slope of -2.}
\end{center}
\end{figure}

We next briefly discuss the two order parameters that were used to help group runs into mode families. The first, $\phi$, was the angle of the direction of maximum extension to the $z$-axis, allowed to vary between 0 and $\pi/2$. $\phi$ was determined by finding the eigenvector of the largest eigenvalue of the radius of gyration tensor. The second, $C_2$, was defined as follows
\begin{equation}
C_2 = \frac{1}{NR}\sum_{i}min(\left|\vec{r}_i-\vec{r}_j^{\: \prime}\right|)
\label{C_2}
\end{equation} 
where $R$ is the average bead separation and $\vec{r}_j^{\: \prime}$ are the bead positions rotated about the $z$-axis by $\pi$ in the centre of mass frame: the minimum distance from each bead to a bead in the rotated configuration is summed. Smaller values of $C_2$ indicate configurations which are closer to being symmetric under a $\pi$ rotation.

Figs.~\ref{fig:N50_modes_sup} (a) and (b) show the values for these two order parameters for different $\alpha$ for the $N = 50$ results. Each point is the average over one of fifty runs -- they are plotted in an arbitrary order. It should be emphasised that all the points within two consecutive vertical lines are for different runs for the same $\alpha$ -- the different positions along the $x$-axis within each section are irrelevant.

We also include a plot of the power-spectrum of the data for the chaotic mode plotted in Fig.~\ref{fig:chaotic_pow_spec}. This was obtained by taking the modulus-squared of the discrete Fourier transform of the displacement of the average bead position around its overall drift. As may be seen from Fig.~\ref{fig:chaotic_pow_spec}, the exponent of the decay is close to -2 (the measured value is $-1.94\pm0.03$).

\end{document}